\documentclass[12pt]{JHEP3}

\usepackage{amsmath}

\newcommand{\be}[1]{ \begin{equation}\label{#1} }
\newcommand{\ee}{\end{equation}}
\newcommand{\ben}[1]{\begin{eqnarray}\label{#1} }
\newcommand{\een}{\end{eqnarray}}

\newcommand{\p}{\partial}
\newcommand{\refb}[1]{(\ref{#1})}

\newcommand{\FF}{{\cal F}}

\newcommand{\DD}{{\cal D}}

\newcommand{\LL}{{\cal L}}

\title{Topologically Massive Higher Spin  Gravity}

\author{
Arjun Bagchi$^1$\footnote{arjun DOT bagchi AT ed DOT ac DOT uk}, Shailesh Lal$^2$\footnote{shailesh AT hri DOT res DOT in}, Arunabha Saha$^2$\footnote{arunabha AT hri DOT res DOT in}, Bindusar Sahoo$^3$\footnote{bsahoo AT ictp DOT it}\\

$^1$ $\,$School of Mathematics, \\
$\;$ $\,$University of Edinburgh \\
$\;$ $\,$Edinburgh EH9 3JZ, UK. \\

$\;$ $^2$Harish-Chandra Research Institute,\\
$\;$ $\,$Chhatnag Road, Jhusi,\\
$\;$ $\,$Allahabad 211019, India.\\

$\;$ $^3$ICTP, High Energy, Cosmology and Astroparticle Physics,\\
$\;$ $\,$Strada Costiera 11, 34151, Trieste, Italy.\\

}

\abstract{We look at the generalisation of topologically massive gravity (TMG) to higher spins, specifically spin-3. We find a special ``chiral" point for the spin-three, analogous to the spin-two example, which actually coincides with the usual spin-two chiral point. But in contrast to usual TMG, there is the presence of a non-trivial trace and its logarithmic partner at the chiral point. The trace modes carry energy opposite in sign to the traceless modes. The logarithmic partner of the traceless mode carries negative energy indicating an instability at the chiral point. We make several comments on the asymptotic symmetry and its possible deformations at this chiral point and speculate on the higher spin generalisation of $LCFT_2$ dual to the spin-3 massive gravity at the chiral point.}

\preprint{EMPG-11-20\\HRI/ST/1109}
\begin{document}

\baselineskip 3.5ex

\section{Introduction}

Gravity in three dimensions has long been a testing ground for constructing a theory of quantum gravity in higher dimensions. Although the actual solutions are quite different from say gravity in four dimensions, the three dimensional models have been instructive for the analysis of more conceptual problems like the role of topology and topology-change, the connections between different quantisation procedures. 
As is well known, the main difference of three dimensional gravity with higher dimensional gravity arises from the fact that there are no local degrees of freedom for gravity in 3d. There are no gravitational waves and curvature is concentrated at the locations of matter. For topologically trivial spacetimes, there are no gravitational degrees of freedom at all. 

To make the dynamics of three dimensional gravity more like gravity in higher dimensions, one needs to restore local degrees of freedom. In 3d, there is the unique opportunity of adding a gravitational Chern-Simons term to the action which now becomes
\ben{}
&& S_{3} = S_{EH} + S_{CS} \\
&\mbox{where}& \quad S_{EH} = \int d^3 x \sqrt{-g} (R - 2\Lambda) \\ \label{gravcs}
&\mbox{and}& \quad S_{CS} =  \frac{1}{2\mu} \int d^3 x  {\epsilon}^{\mu\nu\rho}
\left(\Gamma^\sigma_{\mu\lambda}\partial_\nu\Gamma^\lambda_{\rho\sigma}+\frac{2}{3}\Gamma^\sigma_{\mu\lambda}
\Gamma^\lambda_{\nu\theta}\Gamma^\theta_{\rho\sigma}\right) 
\een
The linearlised equations of motion of this theory are those of a massive scalar field. The existence of this massive excitation can also be traced to the effective interaction of static external sources where one finds a Yukawa attraction with interaction energies as expected for a massive scalar graviton. 
The theory is called topologically massive gravity \cite{Deser:1981wh, Deser:1982vy} 

Topologically massive gravity theories in three dimensions with a negative cosmological constant ($\Lambda = - 1/\ell^2$) have been recently extensively studied in the context of AdS/CFT \cite{Maldacena:1997re}.  Without the Chern-Simons term, 3d gravity in AdS space has the additional feature of having black hole solutions \cite{BTZ}. Now with the topological term, we have both black holes and propagating gravitons. For a generic value of the coefficient of the gravitational Chern-Simons term, the theory has been shown to be inconsistent: either the black hole or the gravitational waves have negative energy. It was conjectured in \cite{Li:2008dq} that the theory becomes sensible at a special point where $\mu \ell =1$. The authors claimed that the dual boundary CFT became a chiral CFT with one of the central charges vanishing ($c_L=0$). This claim, however, was soon hotly contested \cite{confusion} and in following works \cite{Grumiller:2008qz}, topologically massive gravity at the chiral point was shown to be more generally dual to a logarithmic CFT. The energies of these logarithmic solutions were calculated and it was shown that these carried negative energy at the chiral point indicating an instability and the breakdown of the Chiral gravity conjecture. 
A more complete analysis based on techniques of holographic renormalisation showed that this claim was indeed justified \cite{Skenderis:2009nt}. It was discussed that the original chiral gravity conjecture might also hold in a limited sense when one can truncate the LCFT to a chiral CFT provided certain three-point functions vanish {\footnote {The existence of such a truncation only shows that a set of operators of the LCFT form a closed sub-sector, not that this sub-sector has a dual of its own \cite{Skenderis:2009nt}.}}. Similar claims were also made in \cite{Maloney:2009ck}.

Higher-spin theories in $AdS_3$ have been the subject of active interest recently. Unlike their higher-dimensional cousins, they admit a truncation to an arbitrary maximal spin $N$, rather than involving the customary infinite tower of higher-spin fields. Also, like gravity, they possess no propagating degrees of freedom (see, for example, \cite{Campoleoni:2010zq} or \cite{Gaberdiel:2010ar}). The asymptotic symmetry structure for theories with higher spin in AdS have been examined in \cite{Henneaux:2010xg, Campoleoni:2010zq} (see also the recent work \cite{Campoleoni:2011hg}). The authors find that a Brown-Henneaux \cite{BH} like analysis for a theory with maximal spin-$N$ in the bulk yields a $W_N$ asymptotic symmetry algebra. For the spin-3 example, this is the non-linear classical $W_3$ algebra. This has been tested at the one-loop level in \cite{Gaberdiel:2010ar}, using the techniques developed in \cite{David:2009xg}. Finally, this lead to the proposal of a duality between a family of higher-spin theories in $AdS_3$ and $W_N$ minimal models in the large $N$ limit in \cite{Gaberdiel:2010pz}, which has subsequently been checked in \cite{other}. 

Motivated by the features of topologically massive gravity recounted previously, a natural question to ask is what happens when these higher-spin theories are similarly deformed by the addition of a Chern-Simons term. In this paper, we initiate a study of these issues by considering the effect of parity violating, three-derivative terms added to the quadratic action of spin-$3$ Fronsdal fields in $AdS_3$. These are the spin-$3$ analogues of the linearisation of the  gravitational Chern-Simons term described in \refb{gravcs}, and we shall continue to refer to them as ``Chern-Simons" terms.

The outline of the paper is as follows: we start out in Sec.\ref{DD} by constructing the curved space analogue of the action for massive gravity coupled to higher spin modes in \cite{Damour:1987vm}. 
The equations of motion are derived from there. After relating the coefficient of the spin-three ``Chern-Simons" term to the spin-two term in Sec.\ref{relation} by looking at the frame-like formulation, we enter a detailed analysis of the equations of motion in Sec.\ref{eom}. 

Here in Sec.\ref{eom}, following a strategy similar to the spin-two case, we first re-write the equations in terms of three commuting differential operators. At the chiral point, two of these operators become identical indicating an inadequacy of the basis of solutions and thereby necessitating the existence of a logarithmic solution. We solve the equations of motion explicitly. We find that unlike the spin-2 counterpart, the trace of the spin-3 cannot be generically set to zero and will be responsible for giving rise to non-trivial solutions in the bulk which carry a trace, in addition to the traceless mode. We also construct the logarithmic solutions corresponding to both the trace and traceless mode. We compute energies for all the solutions. Away from the chiral point, the massive traceless mode carries negative energy, making this a generalisation of the spin-two example. The novelty in our analysis is the existence of the trace mode. The massive trace mode carries positive energy away from the chiral point and is not a gauge artefact. At the chiral point, both the traceless and the trace mode have zero energy. The logarithmic partner of the trace mode at the chiral point carries positive energy whereas the logarithmic partner of the traceless mode has negative energy indicating an instability similar to the case of the spin-two example. We also show that massless branch solutions, and hence massive branch solutions at the chiral point, can be gauged away by appropriate choice of residual gauge transformation. This along with the fact that left branch and massive branch solutions carry zero energy at the chiral point suggests that these can be regarded as being gauge equivalent to vacuum. But the logarithmic branch solutions are not pure gauge and the negative energy for the logarithmic partner of the traceless mode is a genuine instability in the bulk, similar to the spin-2 example. Apart from all this we find a peculiar ``resonant" behaviour for the trace modes at $\mu\ell={1\over 2}$, which needs some understanding from the CFT perspective.

In Sec.\ref{symm}, we make several comments on the nature of the asymptotic symmetry with the gravitational Chern-Simons term. At the chiral limit, we argue that the natural symmetry algebra to look at is a contraction of the $W_3$ algebra which essentially reduces to the Virasoro algebra. We comment on other possible realisations at this limit. We end in Sec.\ref{conc} with discussions and comments and directions of future work. A couple of appendices list some detailed calculations omitted from the main text. 

\subsubsection*{Note Added:} While this work was being readied for submission, the paper \cite{Chen:2011vp} was posted on the arXiv which has some overlap with the present paper. There are some important differences, however. Unlike in \cite{Chen:2011vp} we find additional physical spin-one modes (the trace of the spin-three field) that need to be accounted for\footnote{We note that similar trace modes were found in the flat-space analysis of Deser and Damour \cite{Damour:1987vm} that we shall shortly come to. These (with an appropriate sign convention for the action) were interpreted as ghost-like excitations. But as we will see later, as per our sign convention of the action (which is required for the positivity of energy of BTZ black holes \cite{Li:2008dq}), these modes carry positive energy and hence cannot be ghost like. On the contrary the traceless modes will carry negative energy and will be ghost-like.}. The analysis of the spin-3 traceless mode is in agreement with \cite{Chen:2011vp}.  In addition, we also construct all logarithmic solutions and compute their energies and have a different proposal for the asymptotic symmetry algebra. 

\section{Spin-$3$ fields in $AdS_3$ with a Chern Simons term}\label{DD}

We begin by reviewing the linearised action for spin-$3$ Fronsdal fields\footnote{We remind the reader that these fields are completely symmetric rank-$3$ tensors. The usual double-tracelessness constraint would not play a role before the introduction of spin-$4$ fields.} with a Chern-Simons term in flat space \cite{Damour:1987vm} (see also the related work \cite{Bergshoeff:2009tb}). The Fronsdal operator $\mathcal F$ for the spin-$3$ field is given by
\begin{equation}
\label{flatfronsdal}
\mathcal{F}_{MNP}\left[\phi\right]=\partial^{2} \phi_{MNP}-\partial_{(M}\partial^A\phi_{NP)A} +\partial_{(M}\partial_{N}\phi_{P)A}\,^A,
\end{equation}
where the brackets denote the sum of the minimal number of terms necessary to have complete symmetrisation in the enclosed indices without any overall normalisation factor. We then define the tensor $G_{MNP}$ by
\begin{equation}
G_{MNP}=\mathcal{F}_{MNP}-\frac{1}{2}\eta_{(MN}\mathcal{F}_{P)A}\,^A.
\end{equation}
It was shown in \cite{Damour:1987vm} that the most general action with up to three derivatives and parity violating terms could be written as
\begin{equation}
\label{minkMHS}
S\left[\phi\right]=\frac{1}{2}\int d^3x \phi^{MNP}G_{MNP}+\frac{1}{2\mu^\prime}\int d^3x \phi^{MNP}\epsilon_{QR(M}\partial^Q G^R\,_{NP)}
\end{equation}
The two terms appearing in this action are each invariant under the gauge transformation
\begin{equation}
\label{flatgauge}
\phi_{MNP}\mapsto \phi_{MNP}+ \partial_{(M}\xi_{NP)},
\end{equation}
where $\xi$ is a traceless symmetric rank two tensor. The first term is just the usual Fronsdal action for massless spin-$3$ fields \cite{Fronsdal:1978rb}, while the second term is the linearised Chern-Simons term. 

In this paper, we will study the covariantisation of this action to $AdS_3$. To do so, we minimally couple the background gravity to the spin-$3$ fluctuation by promoting all partial derivatives to covariant derivatives, and demanding invariance under the gauge transformations\footnote{In going from flat space to $AdS_3$, in addition to changing partial derivatives to covariant derivatives in \refb{flatfronsdal}, we have to multiply the last term by a factor of $1\over 2$ so that we are consistent with our earlier convention of symmetrisation. With partial derivatives, the last term will have a minimum of three terms whereas with covariant derivatives, it will have six terms, because covariant derivatives do not commute.}
\begin{equation}
\label{gaugeads}
\phi_{MNP}\mapsto \phi_{MNP}+\nabla_{(M}\xi_{NP)},
\end{equation}
where $\nabla$ is the covariant derivative defined using the background $AdS_3$ connection. To construct the $AdS$ generalisation of \refb{minkMHS}, it is helpful to recollect what happens in the case where there is no topological term, \textit{i.e.} the covariantisation of the Fronsdal action.
As reviewed for example in \cite{Campoleoni:2010zq}, the Fronsdal operator \refb{flatfronsdal} (defined now with covariant derivatives instead of partial derivates) is no longer invariant under the gauge transformation \refb{gaugeads}, what is invariant (for the spin-$3$ field in $AdS_3$) is the combination \cite{Fronsdal:1978vb}
\begin{equation} \label{frons}
\tilde{\mathcal F}_{MNP}=\mathcal{F}_{MNP}-\frac{2}{\ell^2}g_{(MN}\phi_{P)A}\,^A,
\end{equation}
and if we now define
\begin{equation}
\label{AdSG}
G_{MNP}=\tilde{\mathcal F}_{MNP}-\frac{1}{2}g_{(MN}\tilde{\mathcal F}_{P)A}\,^A,
\end{equation}
the gauge invariant Fronsdal action is given by \cite{Fronsdal:1978vb}
\begin{equation}
S\left[\phi\right]=\frac{1}{2}\int d^3x \sqrt{-g}\phi^{MNP}G_{MNP}.
\end{equation}
It turns out that the case with the Chern-Simons terms is essentially similar. The gauge invariant action is given by
\begin{equation}
\label{adsMHS}
S\left[\phi\right]=\frac{1}{2}\int d^3x \sqrt{-g}\phi^{MNP}G_{MNP}+\frac{1}{2\mu^\prime}\int d^3x \sqrt{-g} \phi^{MNP} \varepsilon_{QR(M}\nabla^Q G^R\,_{NP)},
\end{equation}
where $G_{MNP}$ is now defined through \refb{AdSG}, and
\begin{equation}
\varepsilon^{MNP}\equiv {1\over \sqrt{-g}}\epsilon^{MNP}.
\end{equation}
We remind the reader that $\varepsilon^{MNP}$ is a tensor and all indices are raised and lowered by the background metric. We can write the above action more compactly by defining
\begin{equation}
\hat{\FF}_{MNP}=\tilde{\FF}_{MNP}+\frac{1}{\mu^\prime}\varepsilon_{QR(M}\nabla^Q \tilde{\FF}^R\,_{NP)},
\end{equation}
in terms of which the action becomes
\begin{equation} \label{action}
S\left[\phi\right]=\frac{1}{2}\int d^3x\sqrt{-g} \phi^{MNP}\left(\hat{\FF}_{MNP}-\frac{1}{2}g_{(MN}\hat{\FF}_{P)}\right).
\end{equation}
One may further show that this action gives rise to the equations of motion
\begin{equation}
\label{linearisedEOM}
\DD^{(M)} \tilde{\FF}_{MNP}\equiv\hat{\FF}_{MNP}=\tilde{\FF}_{MNP}+\frac{1}{\mu^\prime}\varepsilon_{QR(M}\nabla^Q\tilde{\FF}^R_{NP)}=0.
\end{equation}
Alternatively, one could have started with constructing the most general parity violating, three derivative equations of motion for $\phi_{MNP}$ in flat space in three dimensions consistent with the gauge invariance \refb{flatgauge}, and attempted a covariantisation to $AdS$. We had initially followed this procedure and obtained identical results. In the above equations, however, the coefficient $\mu^{\prime}$ is arbitrary and is not fixed by the gauge invariant structure. In the next section, we will look at the relation of our action with the $SL(3,R)\times SL(3,R)$ Chern-Simons formulation of spin-3 gravity \cite{Campoleoni:2010zq} with unequal levels and obtain the relation of $\mu^{\prime}$ with the coefficient of gravitational Chern-Simons term $\mu$, given in terms of the left and right levels $a_L$ and $a_R$ as,
\be{cs}
{{a_L-a_R}\over 2}={1\over \mu}.
\ee

\section{Relation with Chern-Simons formulation of high spin gravity and fixing the normalisation}
\label{relation}
It has been observed in \cite{Henneaux:2010xg, Campoleoni:2010zq} that higher spin gravity in three dimensions can have a Chern-Simons formulation. The levels of the Chern-Simons action in \cite{Henneaux:2010xg, Campoleoni:2010zq}, were taken to be equal and hence it produced only the higher-spin extension of Einstein gravity. Since it is known that if we take unequal levels of the Chern-Simons action in pure gravity and impose the torsion constraints, we get parity violating Chern-Simons term and the action becomes that of a topologically massive gravity. We should also be able to do the same for spin-3 massive gravity by taking unequal levels of the Chern-Simons terms. After taking unequal levels for the $SL(3,R)\times SL(3,R)$ Chern-Simons action in \cite{Campoleoni:2010zq}, and imposing the torsion constraints, we arrive at the following action
\ben{7}
S&=&{1\over{8\pi G}}\int e^a \wedge\left(d\omega_{a}+{1\over 2} \epsilon_{abc}\omega^{b}\wedge\omega^{c}-2\sigma\epsilon_{abc}\omega^{bd}\wedge\omega_{d}^{~c}\right) \nonumber \\
&& ~~~~~~~~~ -2\sigma e^{ab}\wedge\left(d\omega_{ab}+2\epsilon_{cda}\omega^{c}\wedge\omega_{b}^{~d}\right)
+{1\over 6l^2}\epsilon_{abc}\left(e^a\wedge e^b \wedge e^c -12\sigma e^a \wedge e^{bd} \wedge e_{d}^{~c}\right) \nonumber \\
&& +{1\over \mu}\int \omega^a \wedge d\omega_{a} +{1\over 3}\epsilon_{abc}\omega^{a}\wedge\omega^{b}\wedge\omega^{c} -2\sigma\omega^{ab}\wedge d\omega_{ab}-4\sigma\epsilon_{abc}\omega^{a}\wedge\omega^{b}_{~d}\wedge \omega^{dc}. 
\een
Subject to the torsion constraint
\ben{1.1}
&& de^a +\epsilon^{abc}\omega_b \wedge e_c -4\sigma\epsilon^{abc}e_{bd}\wedge\omega^{d}_{~c}=0, \nonumber \\
&& de^{ab}+\epsilon^{cd(a}\omega_{c}\wedge e_{d}^{~b)}+\epsilon^{cd(a} e_{c}\wedge \omega_{d}^{~b)}=0.
\een
This is the full non-linear action for spin-3 massive gravity. But since we are interested in linearised equations of motion, we can expand this action around $AdS_3$ background
\be{1.11}
\bar{e}^a=e^{a}_{AdS}, \quad \bar{e}^{ab}=0.
\ee
And then take linearised fluctuations $h_{M}^{~a}$ and $h_{M}^{~ab}$ around this background. And finally we should be able to write everything in terms of the physical Fronsdal fields $\tilde{h}_{MN}$ and $\phi_{MNP}$, defined as
\ben{1.12}
\tilde{h}_{MN}&=& {1\over 2}\bar{e}_{(M}^{~a}h_{N)a}, \nonumber \\
\phi_{MNP} &=& {1\over 3}\bar{e}_{(M}^{a}\bar{e}_{N}^{~b}h_{P)ab}.
\een
The above action \refb{7} is, however, given in terms of the frame fields
\ben{1.13}
h_{MN} &=& \bar{e}_{M}^{~a}h_{N a}, \nonumber \\
h_{MNP} &=& \bar{e}_{M}^{a}\bar{e}_{N}^{~b}h_{Pab}.
\een
The frame fields has an additional $\Lambda$ gauge symmetry \cite{Campoleoni:2010zq} which can be gauge fixed to write down the entire action in terms of the physical Fronsdal fields \refb{1.12}. 

If one is able to successfully implement the programme, one should arrive at the action \refb{action}, since the structure is completely determined by gauge invariance. Since we already have the action, we will bypass the complete programme and just use the Chern-Simons formulation to fix the normalisation of the coefficient $\mu^{\prime}$. For that it is sufficient to find the coefficient of some simple terms. Hence, we use the action \refb{7}, to find the coefficients of $\phi^{MNP}\nabla^2 \phi_{MNP}$ and $\phi^{MNP}\epsilon_{QRM}\nabla^{Q}\nabla^2 \phi_{NP}^{~~R}$. These coefficients can be found after a simple exercise and the quadratic action is
\ben{5.3}
S &=&  {1\over 2} \int \sqrt{-g} \left(\phi^{MNP}\nabla^2 \phi_{MNP} + {1\over 2 \mu} \phi^{MNP}\epsilon_{QRM}\nabla^{Q}\nabla^2 \phi_{NP}^{~~R} +\cdots \right).
\een
Here we have used ${1\over 16\pi G}=1$ and $2\sigma=-1$. Comparing the coefficients of the above terms to the coefficient of similar terms in \refb{action}, we see that, $\mu$ and $\mu^{\prime}$ are related by
\be{5.5}
\mu^{\prime}=6\mu.
\ee
\section{Analysis of the linearised equations of motion}\label{eom}

\subsection{Solving the linearised equations of motion}

In this section, we will analyse the linearised equations of motion \refb{linearisedEOM}. We wish to cast this equation in a form $\DD^{(M)}\DD^{(L)}\DD^{(R)}\phi_{MNP}=0$ for three commuting differential operators $\DD^{(M)}$, $\DD^{(L)}$ and $\DD^{(R)}$. $\DD^{(M)}$ is defined in  \refb{linearisedEOM}. So we have to put $\tilde{\FF}_{MNP}$ \refb{frons} into the form $\DD^{(L)}\DD^{(R)}\phi_{MNP}$. Note that generically this cannot be done. One has to do a suitable field redefinition and use a suitable gauge condition to be able to do it. After a careful analysis, one finds that there is a unique field redefinition and gauge condition which solves the above purpose. They are
\ben{4.3}
\phi_{MNP} &=& \tilde{\phi}_{MNP}-{1\over 9}g_{(MN}\tilde{\phi}_{P)}, \nonumber \\
\nabla^{Q}\tilde{\phi}_{QMN} &=& {1\over 2}\nabla_{(M}\tilde{\phi}_{N)}.
\een
Using this field redefinition and gauge condition, we get
\be{4.5}
\tilde{\FF}_{MNP}=\nabla^2 \tilde{\phi}_{MNP}-{1\over 6}\nabla_{(M}\nabla_{N}\tilde{\phi}_{P)}-{8\over 9l^2}g_{(MN}\tilde{\phi}_{P)}-{1\over 9}\nabla^2 \tilde{\phi}_{(M}g_{NP)}+{1\over 9}g_{(MN}\nabla_{P)}\nabla^Q \tilde{\phi}_Q.
\ee
One can further see that this $\tilde{\FF}_{MNP}$ can be cast into the desired form as
\be{4.6}
\tilde{\FF}_{MNP}=-{4\over \ell^2}\DD^{(R)}\DD^{(L)}\tilde{\phi}_{MNP},
\ee
where $\DD^{(R)}$ and $\DD^{(L)}$ are defined as
\ben{4.7}
\DD^{(L)}\tilde{\phi}_{MNP} &=& \tilde{\phi}_{MNP}+{\ell\over 6} \varepsilon^{QR}_{~~~(M|}\nabla_{Q}\tilde{\phi}_{R|NP)}, \nonumber \\
\DD^{(R)}\phi_{MNP} &=& \tilde{\phi}_{MNP}-{\ell\over 6} \varepsilon^{QR}_{~~~(M|}\nabla_{Q}\tilde{\phi}_{R|NP)}.
\een
Now, putting this together with \refb{linearisedEOM}, our equations of motion become
\be{4.8}
\DD^{(M)}\DD^{(L)}\DD^{(R)}\tilde{\phi}_{MNP}=0.
\ee
One can also check that $\DD^{(M)}$, $\DD^{(L)}$ and $\DD^{(R)}$ are three sets of mutually commuting operators. The superscripts $(M)$, $(L)$ and $(R)$ stand for massive, left moving and right moving branches, respectively.
Taking trace of the equation \refb{4.8} and contracting it with $\nabla^M$, one finds that
\be{4.9}
\nabla^{M}\tilde{\phi}_{M}=0
\ee
However, we see that we do not get any tracelessness constraint from the equation of motion and we will soon see that the trace will be responsible for giving rise to some non-trivial solutions to the equation of motion.

Let us now try to solve for the massive branch. We can obtain the left moving and right moving solution from this by putting $\mu\ell=1$ and $\mu\ell=-1$ respectively. The massive branch equation is
\be{6}
\DD^{(M)}\tilde{\phi}_{MNP}=0,
\ee
where $\DD^{(M)}$ is defined in \refb{linearisedEOM}. Let $\tilde{\DD}^{(M)}$ be the same as $\DD^{(M)}$ with $\mu \to -\mu$. By acting on \refb{6} with $\tilde{\DD}^{(M)}$, we get
\ben{6.1}
\nabla^2 \tilde{\phi}_{MNP}&-&\left(4\mu^2 -{4\over \ell^2}\right)\tilde{\phi}_{MNP} \nonumber \\
&=& {1\over 6}\nabla_{(M}\nabla_{N}\tilde{\phi}_{P)} + {8\over 9 \ell^2}g_{(MN}\tilde{\phi}_{P)}+{1\over 9}\nabla^2 \tilde{\phi}_{(M}g_{NP)}.
\een
The equations for the massless branch is the same as above with $\mu \to {1\over \ell}$. Taking the trace of the above equation, we get
\be{6.2}
\left(\nabla^2 -36 \mu^2 + {2\over \ell^2}\right)\tilde{\phi}_M=0.
\ee
We will solve the equations in $AdS_3$ background with the metric
\be{ADS}
ds^2=\ell^2\left(-\cosh^{2}\rho d\tau^2+\sinh^{2}\rho d\phi^2+d\rho^2\right).
\ee
The metric has the isometry group $SL(2,R)_{L}\times SL(2,R)_{R}$. The $SL(2,R)_{L}$ isometry generators are \cite{Li:2008dq}
\ben{isom}
L_0 &=& i\p_{u}, \nonumber \\
L_{-1} &=& ie^{-iu}\left[{{\cosh 2\rho}\over {\sinh 2\rho}} \p_u - {1\over {\sinh 2\rho}} \p_v +{i\over 2} \p_{\rho}\right], \nonumber \\
L_{1} &=& ie^{iu}\left[{{\cosh 2\rho}\over {\sinh 2\rho}} \p_u - {1\over {\sinh 2\rho}} \p_v -{i\over 2} \p_{\rho}\right],
\een
where $u\equiv \tau+\phi$ and $v\equiv \tau-\phi$. The $SL(2,R)_{R}$ generators $(\bar{L}_0,\bar{L}_{\pm1})$ are given by the above expressions with $u\to v$. The quadratic Casimirs are
\ben{cas}
L^2 &=& {1\over 2}\left(L_{1}L_{-1}+L_{-1}L_{1}\right)-L_{0}^{2}, \nonumber \\
\bar{L}^2 &=& {1\over 2}\left(\bar{L}_{1}\bar{L}_{-1}+\bar{L}_{-1}\bar{L}_{1}\right)-\bar{L}_{0}^{2}.
\een
The Laplacian acting on tensors of various ranks can be written in terms of $SL(2,R)$ Casmirs as
\ben{6.3}
\nabla^2 h &=& -{2\over \ell^2}\left(L^2 +\bar{L}^2\right)h, \nonumber \\
\nabla^2 h_M &=& -{2\over \ell^2}\left(L^2 +\bar{L}^2\right)h_M -{2\over \ell^2}h_M, \nonumber \\
\nabla^2 h_{MN} &=&  -{2\over \ell^2}\left(L^2 +\bar{L}^2\right)h_{MN} -{6\over \ell^2}h_{MN} +{2\over \ell^2}h g_{MN}, \nonumber \\
\nabla^2 h_{MNP} &=&  -{2\over \ell^2}\left(L^2 +\bar{L}^2\right)h_{MNP} -{12\over \ell^2}h_{MNP} +{2\over \ell^2}h_{(M} g_{NP)}. 
\een
Now we are in a position to solve the equations of motion. We will first solve for the trace \refb{6.2}, put it back into the full equation \refb{6.1} and obtain the solution to the full equation which carries this trace.
Using \refb{6.3}, we can solve for the trace and classify it in terms of $SL(2,R)$ primaries and descendants. Using \refb{6.3}, we can write \refb{6.2} as
\be{6.4}
\left[-2\left(L^2+\bar{L}^2\right)-36\mu^2 \ell^2\right]\tilde{\phi}_{M}=0.
\ee
Let us specialise to ``primary" states with weights $(h,\bar{h})$, i.e
\ben{6.5}
 && L_{0} \tilde{\phi}_M = h \tilde{\phi}_M, \quad \bar{L}_{0} \tilde{\phi}_M = \bar{h} \tilde{\phi}_M,  \nonumber \\
&&  L_{1} \tilde{\phi}_M = 0, \quad \bar{L}_{1} \tilde{\phi}_M = 0.
\een
From the explicit form of the generators \refb{isom}, one can see that $(u,v)$ dependence of $\tilde{\phi}_{M}$ is
\be{6.6}
 \tilde{\phi}_M=e^{-ihu -i\bar{h}v}\psi_{M}(\rho),
\ee
The primary conditions ( second line of \refb{6.5}) are satisfied for $h-\bar{h}=0, \pm1$, but the only solutions compatible with the condition $\nabla^{M}\tilde{\phi}_{M}=0$ are
\ben{6.8}
&& h-\bar{h}=1, \quad \psi_{v}=0, \quad \psi_{\rho}={2i\over {\sinh}(2\rho)}f(\rho) , \quad \psi_{u}=f(\rho), \nonumber \\
\texttt{or} && h-\bar{h}=-1, \quad \psi_{u}=0, \quad \psi_{\rho}={2i\over {\sinh}(2\rho)}f(\rho) , \quad \psi_{v}=f(\rho), \nonumber \\
\een
where $f(\rho)$ satisfies \footnote{We have put an overall factor of $1\over \ell^2$ in the solution to $f(\rho)$. This is because (for dimensional consistency) we want to obtain the solution to $\tilde{\phi}_{MNP}$ which are dimensionless so that at the end of the day we can multiply appropriate powers of $\ell$ to the solution to match it with its canonical dimension. And since we want the full solution to be dimensionless, the trace has to be multiplied by the factor of $1\over \ell^2$}
\ben{6.9}
&& \p_{\rho} f(\rho)+\left[{{(h+\bar{h})\sinh^{2}(\rho)-\cosh^{2}(\rho)}\over {\sinh \rho \cosh \rho}}\right]f(\rho)=0 \nonumber \\
&& \implies f(\rho)={1\over \ell^2}\left(\cosh \rho\right)^{-(h+\bar{h})}\sinh(\rho).
\een
The first line of \refb{6.8} is the solution to our original equation of motion \refb{6}, whereas the second line is the solution to the original equation of motion with $\mu\to -\mu$. The second line will therefore not belong to the massive branch, but by putting $\mu\ell=1$ in the second line we will get the right branch solution and by putting $\mu\ell=1$ in the first line, we will get the left branch solution. Putting \refb{6.8} in \refb{6.2}, we get
\ben{6.10}
&& h=1 \pm 3\mu\ell, \quad \bar{h}=\pm 3\mu\ell, \nonumber \\
\texttt{or} && h=\pm 3\mu \ell  \quad \bar{h}=1 \pm 3\mu\ell.
\een
It is easy to see that $f(\rho)$ in \refb{6.9} will blow up at $\rho \to \infty$ if $h+\bar{h} <1$. Since $\mu\ell \geq 1$, this rules out the lower sign in \refb{6.10}. To summarise, the different branch solution will carry the following weights.
\ben{6.101}
\texttt{Massive:} \quad & h=1+3\mu\ell \quad & \bar{h}=3\mu\ell, \nonumber \\
\texttt{Left:} \quad & h=4 \quad & \bar{h}=3, \nonumber \\
\texttt{Right:} \quad & h=3 \quad & \bar{h}=4.
\een
We can successively apply $L_{-1}$ and $\bar{L}_{-1}$ on the primary solutions obtained above and obtain the descendant solutions. After obtaining the solution for the trace, let us try to obtain the solution to the full equation \refb{6.1}. Using \refb{6.3}, we can write \refb{6.1} as
\be{6.11}
{1\over\ell^2}\left[-2\left(L^2 +\bar{L}^2\right)-8-4\mu^2 \ell^2\right]\tilde{\phi}_{MNP}={1\over 6}\nabla_{(M}\nabla_{N}\tilde{\phi}_{P)}-{4\over 3\ell^2}\left(1-3\mu^2 \ell^2\right)\tilde{\phi}_{(M}g_{NP)}.
\ee
We have to put the solution obtained for the trace in the RHS of the above equation and obtain the solution to the full equation. If we take the primary (or descendant) trace solutions (\ref{6.6},\ref{6.8},\ref{6.9}) in the RHS of \refb{6.11}, then one can show that $\tilde{\phi}_{MNP}$, should also be a primary (or descendant) solution. This is because of the following identity (which we prove in appendix \ref{iso})
\be{isom}
L_{\xi} \nabla_{(M}\nabla_{N}\tilde{\phi}_{P)}=  \nabla_{(M}\nabla_{N}L_{\xi}\tilde{\phi}_{P)},
\ee
where $L_{\xi}$ is an isometry generator.

Since the trace carries weights $(h,\bar{h})$ given by \refb{6.10}, we can break the full $\tilde{\phi}_{MNP}$ as
\be{6.111} 
\tilde{\phi}_{MNP}=\chi_{MNP}+\Sigma_{MNP},
\ee
where all the parts of $\tilde{\phi}_{MNP}$ which carry the weights $(h,\bar{h})$ are put into $\chi_{MNP}$ and the rest in $\Sigma_{MNP}$. They satisfy the equations
\ben{6.12}
{1\over\ell^2}\left[-2\left(L^2 +\bar{L}^2\right)-8-4\mu^2 \ell^2\right]{\chi}_{MNP} &=& {1\over 6}\nabla_{(M}\nabla_{N}\tilde{\phi}_{P)}-{4\over 3 \ell^2}\left(1-3\mu^2 \ell^2\right)\tilde{\phi}_{(M}g_{NP)}, \nonumber \\
{1\over \ell^2}\left[-2\left(L^2 +\bar{L}^2\right)-8-4\mu^2 \ell^2\right]{\Sigma}_{MNP} &=& 0.
\een
Since the RHS of \refb{6.11} carries the weights \refb{6.10}, hence it should be equated with a part of LHS which carries the same weights and hence the equation is decomposed in the above way. The first of the equation in \refb{6.12} becomes (by using the weights \refb{6.10})
\be{6.13}
{8\over \ell^2}\left(4\mu^2 \ell^2-1\right)\chi_{MNP}={1\over 6}\nabla_{(M}\nabla_{N}\tilde{\phi}_{P)}-{4\over 3 \ell^2}\left(1-3\mu^2 \ell^2\right)\tilde{\phi}_{(M}g_{NP)}.
\ee
The solution to $\chi_{MNP}$ is therefore
\be{6.14}
\chi_{MNP}={\ell^2\over 8 \left(4\mu^2 \ell^2-1\right)}\left[{1\over 6}\nabla_{(M}\nabla_{N}\tilde{\phi}_{P)}-{4\over 3 \ell^2}\left(1-3\mu^2 \ell^2\right)\tilde{\phi}_{(M}g_{NP)}\right].
\ee
We see that the solution has a divergence at $\mu\ell={1\over 2}$. This is not something unusual since we are solving the equation with a source (RHS of \refb{6.11}) of specific weights $(h,\bar{h})$. This divergent behaviour is analogous to the resonance in forced oscillations. From \refb{6.14}, we notice that
\be{6.141}
g^{NP}\chi_{MNP}=\tilde{\phi}_{M} \quad \nabla^{M}\chi_{MNP}={1\over 2} \nabla_{(N}\tilde{\phi}_{P)}.
\ee
Using \refb{6.141} in the decomposition \refb{6.111} and in the gauge condition \refb{4.3}, we get
\be{6.15}
g^{NP}\Sigma_{MNP}=0, \quad \nabla^{M}\Sigma_{MNP}=0.
\ee
Let us now solve the equation of motion for $\Sigma_{MNP}$ (the second line of \refb{6.12}) subject to the tracelessness and gauge condition \refb{6.15} \footnote{This solution is similar to the one obtained in \cite{Chen:2011vp}.}. We specialise to ``primary" states with weights $(h,\bar{h})$, i.e
\ben{6.19}
 && L_{0} {\Sigma}_{MNP} = h {\Sigma}_{MNP}, \quad \bar{L}_{0} {\Sigma}_{MNP} = \bar{h} {\Sigma}_{MNP}  \nonumber \\
&& L_{1} {\Sigma}_{MNP} = 0, \quad \bar{L}_{1} {\Sigma}_{MNP} = 0.
\een
From the explicit form of the generators, one can see that the $(u,v)$ dependence of $\Sigma_{MNP}$ is
\be{6.20}
 {\Sigma}_{MNP}=e^{-ihu -i\bar{h}v}\sigma_{MNP}(\rho),
\ee
The primary conditions are solved for $h-\bar{h}=0, \pm 1, \pm 2, \pm 3$. But the only solutions compatible with the gauge conditions and tracelessness condition \refb{6.15} are
\ben{6.22}
 h-\bar{h}&=&3, \nonumber \\
 \sigma_{MN v}&=&0 \nonumber \\
 \sigma_{\rho u u}={i f(\rho) \over {\cosh{\rho} \sinh{\rho}}} \quad \sigma_{uuu}=f(\rho) \quad \sigma_{\rho\rho\rho}&=&{-i f(\rho) \over {\cosh^{3}({\rho}) \sinh^{3}({\rho})}} \quad \sigma_{u \rho \rho}= {- f(\rho) \over {\cosh^{2}({\rho}) \sinh^{2}({\rho})}},\nonumber \\
\een
and
\ben{6.23}
 h-\bar{h}&=& -3, \nonumber \\
\sigma_{MN u}&=&0 \nonumber \\
\sigma_{\rho vv}={i f(\rho) \over {\cosh{\rho} \sinh{\rho}}} \quad \sigma_{vvv}=f(\rho) \quad \sigma_{\rho\rho\rho}&=&{-i f(\rho) \over {\cosh^{3}({\rho}) \sinh^{3}({\rho})}} \quad \sigma_{v \rho \rho}= {- f(\rho) \over {\cosh^{2}({\rho}) \sinh^{2}({\rho})}},\nonumber \\
\een
where $f(\rho)$ satisfies
\ben{6.24}
&& \p_{\rho} f(\rho)+\left[{{(h+\bar{h})\sinh^{2}(\rho)-3\cosh^{2}(\rho)}\over {\sinh \rho \cosh \rho}}\right]f(\rho)=0 \nonumber \\
&& \implies f(\rho)=\left(\cosh \rho\right)^{-(h+\bar{h})}\sinh^{3}(\rho).
\een
Now putting the above into the second line of \refb{6.12}, we get
\ben{6.25}
&& h=2 \pm \mu\ell \quad \bar{h}=-1 \pm \mu\ell \nonumber \\
\texttt{or} && h= -1 \pm \mu\ell \quad \bar{h}= 2 \pm \mu\ell
\een
The solution with $h-\bar{h}=3$ belongs to the original massive branch whereas $h-\bar{h}=-3$ belongs to the massive branch with $\mu \to -\mu$. The left branch is obtained by putting $\mu\ell=1$ in the $h-\bar{h}=3$ solution and right branch is obtained by putting $\mu\ell=1$ in the $h-\bar{h}=-3$ solution. It is also easy to check that $f(\rho)$ in \refb{6.24} diverges at $\rho \to \infty$ unless $h+\bar{h}\geq 3$. This rules out the lower sign in \refb{6.25}. To summarise we obtain the following solution
\ben{6.26}
\texttt{Massive:} \quad & h=2+\mu\ell \quad & \bar{h}=-1+\mu\ell \nonumber \\
\texttt{Left:} \quad & h=3 & \bar{h}=0 \nonumber \\
\texttt{Right:} \quad & h=0 \quad & \bar{h}=3
\een
We can successively apply $L_{-1}$ and $\bar{L}_{-1}$ on the primary solutions obtained above to obtain the descendant solutions. At the chiral point $\mu\ell=1$, the massive and left branch solutions coincide and and hence the basis of solutions become insufficient to describe the dynamics. However following the construction of \cite{Grumiller:2008qz}, one sees that a new logarithmic mode emerges (which is annihilated by $\DD^{(L)2}$ and not by $\DD^{(L)}$). We now turn to this point.
\subsection{Logarithmic modes at the chiral point}
Let us denote the massive branch, left branch and right branch solutions with superscripts M, L and R respectively. At the chiral point $\mu\ell=1$, the massive branch and left branch coincides and hence the basis of solutions become insufficient to describe the dynamics. However following the construction of \cite{Grumiller:2008qz}, one sees that a new logarithmic mode emerges (which is annihilated by $\DD^{(L)2}$ and not by $\DD^{(L)}$). The logarithmic mode is obtained as
\be{l1}
   {\Phi}^{(new)}=\lim_{\mu\ell \to 1}{{\Phi^{(M)}(\mu\ell)-\Phi^{(L)}}\over {\mu\ell-1}}={d\Phi^{(M)}(\epsilon)\over d \epsilon}|_{\epsilon=0},
\ee
where $\epsilon \equiv \mu\ell -1$. We have schematically used $\Phi$ to denote any mode which has a decomposition into massless and massive branches and have suppressed any possible spacetime indices. It can be easily seen that since $\Phi^{(M)}$ and $\Phi^{(L)}$ are annihilated by $\DD^{(M)}$ and $\DD^{(L)}$ respectively, the term inside the limit is annihilated by $\DD^{(M)}\DD^{(L)}$ but not by $\DD^{(M)}$ or $\DD^{(L)}$ separately. After taking the limit, therefore the mode is annihilated by $\DD^{(L)2}$ but not by $\DD^{(L)}$. Now let us find out the logarithmic partner of the mode $\chi_{MNP}$ in \refb{6.14}. Expressing $\mu\ell$ in terms of $\epsilon$ and then taking the derivative wrt $\epsilon$, we get
\ben{l2}
\hat{\chi}_{MNP} & \equiv & {{d\chi_{MNP}(\epsilon)}\over {d \epsilon}}|_{\epsilon=0} \nonumber \\
&=& -{\ell^2\over 9}\left[{1\over 6}\nabla_{(M}\nabla_{N}\tilde{\phi}_{P)}^{(L)}-{1\over 3 \ell^2}\tilde{\phi}_{(M}^{(L)}g_{NP)}\right] +{\ell^2 \over 24}\left[{1\over 6}\nabla_{(M}\nabla_{N}\hat{\phi}_{P)}+{8\over 3 \ell^2}\hat{\phi}_{(M}g_{NP)}\right], \nonumber \\
\een
where $\tilde{\phi}^{(L)}_{M}$ is the trace of the left branch solution and $\hat{\phi}_{M}\equiv {{d\tilde{\phi}_{M}^{(M)}(\epsilon)}\over {d\epsilon}}|_{\epsilon=0}$. It can be easily seen from the definition of $\hat{\phi}_{M}$ that
\be{l3}
\hat{\phi}_{M}=\left[-3i(u+v)-6\log \cosh \rho\right]\tilde{\phi}_{M}^{(L)},
\ee
and hence
\ben{l4}
&& L_0 \hat{\phi}_{M}= 3\tilde{\phi}_{M}^{(L)}+4\hat{\phi}_{M} \quad {\bar{L}}_0 \hat{\phi}_{M}= 3\tilde{\phi}_{M}^{(L)}+3\hat{\phi}_{M} \quad L_1 \hat{\phi}_{M}=\bar{L}_{1} \hat{\phi}_{M}=0 \nonumber \\
\implies && L^2 \hat{\phi}_{M}=-21\tilde{\phi}_{M}^{(L)}-12\hat{\phi}_{M} \quad \bar{L}^2 \hat{\phi}_{M}=-15\tilde{\phi}_{M}^{(L)}-6\hat{\phi}_{M} \nonumber \\
\implies && \left(\nabla^2 -{34\over \ell^2}\right) \hat{\phi}_M=\left[-{2\over \ell^2}\left(L^2 +\bar{L}^2\right)-{36\over \ell^2}\right]\hat{\phi}_M = {72\over \ell^2}\tilde{\phi}_{M}^{(L)}.
\een
Using the above set of equations and taking the trace of \refb{l2}, we get, as expected, that $\hat{\phi}_{M}$ is the trace of $\hat{\chi}_{MNP}$. We also see that $\hat{\chi}_{MNP}$ satisfies
\be{l6}
L_0 \hat{\chi}_{MNP}= 3{\chi}_{MNP}^{(L)}+4\hat{\chi}_{MNP}, \quad {\bar{L}}_0 \hat{\chi}_{MNP}= 3{\chi}_{MNP}^{(L)}+3\hat{\chi}_{MNP}, \quad L_1 \hat{\chi}_{MNP}=\bar{L}_{1} \hat{\chi}_{MNP}=0.
\ee
We have thus obtained the logarithmic partner of the mode $\chi_{MNP}^{(L)}$ at the chiral point. Using the same trick we can also obtain the logarithmic partner of the mode $\Sigma_{MNP}^{(L)}$ and we get\footnote{This is the same as the logarithmic mode obtained in \cite{Chen:2011vp}.} 
\be{l7}
\hat{\Sigma}_{MNP}\equiv {{d\Sigma_{MNP}^{(M)}(\epsilon)}\over{d\epsilon}}|_{\epsilon=0}=\left[-i(u+v)-2\log\cosh \rho\right]\Sigma_{MNP}^{(L)},
\ee
and hence $\hat{\Sigma}_{MNP}$ satisfies
\be{l8}
L_0 \hat{\Sigma}_{MNP}= {\Sigma}_{MNP}^{(L)}+3\hat{\Sigma}_{MNP} \quad {\bar{L}}_0 \hat{\Sigma}_{MNP}= {\Sigma}_{MNP}^{(L)} \quad L_1 \hat{\Sigma}_{MNP}=\bar{L}_{1} \hat{\Sigma}_{MNP}=0.
\ee

We have so far obtained traceless as well as traceful solutions to the equation of motion \refb{linearisedEOM}. We also obtained their logarithmic partners at the chiral point. We label the massive, left and right branch $\chi$ modes \refb{6.14} as $(M_{\chi})$, $(L_{\chi})$ and $(R_{\chi})$ respectively. We also label the logarithmic solution to the $\chi$ mode \refb{l2} as $({\log}_{\chi})$. Similarly we label the massive, left, right and logarithmic $\Sigma$ modes (\ref{6.22}, \ref{6.23}, \ref{6.24}, \ref{6.26}, \ref{l7}) as $(M_{\Sigma})$, $(L_{\Sigma})$, $(R_{\Sigma})$ and $({\log}_{\Sigma})$ respectively. We will now obtain the energies of all the above modes.

\subsection{Energy of the fluctuations}
After imposing the field redefinition and gauge condition \refb{4.3}, we obtain the action \refb{action} (up to total derivatives) as,
\ben{E1}
S = {1\over 2}\int \sqrt{-g} && \left[-\nabla_{Q}\tilde{\phi}^{MNP}\nabla^{Q}\tilde{\phi}_{MNP}-{1\over 2\mu}\varepsilon_{QRM} \nabla^{Q}\tilde{\phi}^{MNP} \nabla^2 \tilde{\phi}^{R}_{~NP}\right. \nonumber \\
&&+{19\over 9l^2}\left(\tilde{\phi}^{M}\tilde{\phi}_{M} +{1\over 6 \mu}\varepsilon_{QRM}\tilde{\phi}^{M}\nabla^{Q}\tilde{\phi}^{R}\right) \nonumber \\
&&+{17\over 18}\left. \left(\nabla^{Q}\tilde{\phi}^{M}\nabla_{Q}\tilde{\phi}_{M}+{1\over 6\mu}\varepsilon_{QRM}\nabla^{Q}\tilde{\phi}^{M}\nabla^2 \tilde{\phi}^{R}\right)\right]
\een
The momentum conjugate to $\tilde{\phi}_{MNP}$ is 
\ben{E2}
\Pi^{(1)MNP} &\equiv&  {{\delta S}\over{\delta \dot{\tilde{\phi}}_{MNP}}} \nonumber \\
 &=& {\sqrt{-g}\over 2} \left[-\nabla^{0}\left(2\tilde{\phi}_{MNP}+{1\over 6\mu}\varepsilon^{QR(M}\nabla_{Q}\tilde{\phi}_{R}^{~NP)}\right)\right. \nonumber \\
&& +{17\over 18 \times 3}\nabla^{0}\left(2\tilde{\phi}^{(M}g^{NP)}+{1\over 6 \mu}\varepsilon^{QR(M}\nabla_{Q}\tilde{\phi}_{R}g^{NP)}\right) \nonumber \\
&& -\left. {1\over 6\mu}\varepsilon^{0R(M}\nabla^2 \tilde{\phi}_{R}^{~NP)}-{19\over 9\times 18}{1\over \mu \ell^2}\varepsilon^{0R(M}\tilde{\phi}_{R}g^{NP)} +{17\over 18\times 18 \mu}\varepsilon^{0R(M}\nabla^2\tilde{\phi}_{R}g^{NP)}\right]. \nonumber \\
\een
Since we have three time derivatives, we should also implement the Ostrogradsky method (following \cite{Li:2008dq}), and introduce $K_{MNP}\equiv \nabla_{0}\tilde{\phi}_{MNP}$ as a canonical variable and find the momentum conjugate to that which is,
\ben{E3}
\Pi^{(2)MNP} &\equiv&  {{\delta S} \over {\delta \dot{K}_{MNP}}} \nonumber \\
&=& {\sqrt{-g}\over2} \left[{1\over 6\mu}g^{00}\varepsilon^{QR(M}\nabla_{Q}\tilde{\phi}_{R}^{~NP)}-{17\over 18\times 18\mu}g^{00}\varepsilon^{QR(M}\nabla_{Q}\tilde{\phi}_{R}g^{NP)}\right]
\een
The above expressions are the most generic expressions for the conjugate momenta and can be applied on any modes. The conjugate momenta for the different modes are listed in appendix \ref{conjug}. In oder to obtain the energy we must put the expressions for the conjugate momenta in the Hamiltonian
\ben{E31}
H &=& \int d^2 x \left(\dot{\tilde{\phi}}_{MNP}\Pi^{(1)MNP}+\dot{K}_{MNP}\Pi^{(2)MNP}-\LL\right) \nonumber \\
&=&  \int d^2 x \left(\dot{\tilde{\phi}}_{MNP}\Pi^{(1)MNP}-{K}_{MNP}\dot{\Pi}^{(2)MNP}-\LL\right) +{d\over d\tau}\int~ d^2 x ~ {K}_{MNP}{\Pi}^{(2)MNP} \nonumber \\
& \equiv & E^0 + E^1,
\een
where the integral is over $\phi$ and $\rho$ and $\LL$ is the Lagrangian density. We have defined the first integral in the second line of \refb{E31} as $E^0$ and second integral as $E^{1}$. Also note that $\LL=0$ on the solutions. Now we can put the conjugate momenta obtained in appendix \refb{conjug} and the real part of the solutions obtained in the previous sections to get the energy expressions for different modes. One can see by explicitly putting the solutions in the above integrals that $E^{1}$ for all the non-logarithmic modes vanishes but logarithmic modes get non-trivial contribution from $E^{1}$. Putting the real part of the logarithmic solutions and expressions for the conjugate momenta for the logarithmic modes in Mathematica, we get \footnote{All the expressions of energy that we will obtain will have the dimension of $1\over \ell^5$. This is due to our choice of units ${1\over 16\pi G}=1$ and using dimensionless solutions of $\tilde{\phi}_{MNP}$. If we re-instate the factor of ${1\over 16\pi G}=1$ and multiply the solutions of $\tilde{\phi}_{MNP}$ with appropriate powers of $\ell$ matching their canonical dimensions, we will get the correct dimensions of energy. However this will not change any of the qualitative features of the discussion}
\ben{E32}
E^{1}_{({\log}_{\chi})} &=& {d\over d\tau}\int d^2 x ~ {\sqrt{-g}\over 2}\left[-\nabla^{0}\hat{\chi}_{MNP}\left(\hat{\chi}^{MNP}+\chi^{(L)MNP}\right)+{17\over 18}\nabla^{0}\hat{\chi}_{M}\left(\chi^{M}+\chi^{M}\right)
\right], \nonumber \\
&=& {79 \pi \over 280 \ell^{5}} \nonumber \\
E^{1}_{({\log}_{\Sigma})} &=& {d\over d\tau}\int d^2 x ~ {-\sqrt{-g}\over 2}\left[\nabla^{0}\hat{\Sigma}_{MNP}\left(\hat{\Sigma}^{MNP}+\Sigma^{(L)MNP}\right)
\right] \nonumber \\
&=& -{4\pi\over 15 \ell^5}.
\een
We can now put the expressions for the real part of the solutions obtained in the previous sections and conjugate momenta in appendix \refb{conjug}, to get the expressions for $E^{0}$ for different modes. For the non logarithmic $\chi$ modes we get,
\ben{E6}
E^{0}_{(M_{\chi})}&=& -{3\over \mu}\left(3\mu^2-{1\over \ell^2}\right)\int d^2 x ~ \sqrt{-g}~ \varepsilon^{0RM}~ \dot{\chi}_{MNP}^{(M)}\chi_{R}^{(M)~NP} \nonumber \\
&& +{1\over 6\mu}\left(17\mu^2-{5\over \ell^2}\right) \int d^2 x ~ \sqrt{-g}~ \varepsilon^{0RM}~ \dot{\chi}_{M}^{(M)}\chi_{R}^{(M)} \nonumber \\
E^{0}_{(L_{\chi})}&=& \left(-1+{1\over \mu\ell}\right)\int d^2 x ~ \sqrt{-g}~ \left[\dot{\chi}_{MNP}^{(L)}\nabla^{0}\chi^{(L)MNP}-{17\over 18}\dot{\chi}_{M}^{(L)}\nabla^{0}\chi^{(L)M}\right] \nonumber \\
&& -{6\over \mu \ell^2}\int d^2 x ~ \sqrt{-g}~ \varepsilon^{0RM}~ \dot{\chi}_{MNP}^{(L)}\chi_{R}^{(L)~NP} +{2\over \mu \ell^2}\int d^2 x ~ \sqrt{-g}~ \varepsilon^{0RM}~ \dot{\chi}_{M}^{(L)}\chi_{R}^{(L)} \nonumber \\
E^{0}_{(R_{\chi})}&=& \left(-1-{1\over \mu\ell}\right)\int d^2 x ~ \sqrt{-g}~ \left[\dot{\chi}_{MNP}^{(R)}\nabla^{0}\chi^{(R)MNP}-{17\over 18}\dot{\chi}_{M}^{(R)}\nabla^{0}\chi^{(R)M}\right] \nonumber \\
&& -{6\over \mu \ell^2}\int d^2 x ~ \sqrt{-g}~ \varepsilon^{0RM}\dot{\chi}_{MNP}^{(R)}\chi_{R}^{(R)~NP} +{2\over \mu \ell^2}\int\sqrt{-g}\varepsilon^{0RM}\dot{\chi}_{M}^{(R)}\chi_{R}^{(R)} \nonumber \\
\een
For the non logarithmic $\Sigma$ modes, we get
\ben{E61}
E^{0}_{(M_{\Sigma})}&=& {1\over \mu}\left(\mu^2-{1\over \ell^2}\right)\int d^2 x ~ \sqrt{-g}~  \varepsilon^{R0M}~ \dot{\Sigma}_{MNP}^{(M)}\Sigma_{R}^{(M)NP} \nonumber \\
E^{0}_{(L_{\Sigma})}&=& \left(-1+{1\over \mu\ell}\right)\int d^2 x ~ \sqrt{-g}~  \dot{\Sigma}_{MNP}^{(L)}\nabla^{0}\Sigma^{(L)MNP} \nonumber \\
E^{0}_{(R_{\Sigma})}&=& \left(-1-{1\over \mu\ell}\right)\int d^2 x ~ \sqrt{-g}~ \dot{\Sigma}_{MNP}^{(R)}\nabla^{0}\Sigma^{(R)MNP} \nonumber \\
\een
For the logarithmic modes (trace as well as traceless), we get
\ben{E62}
E^{0}_{({\log}_{\chi})}&=& \int d^2 x ~  {\sqrt{-g}\over 2}~ \left[\dot{\hat{\chi}}_{MNP}\nabla^{0}\chi^{(L)MNP}+ \dot{{\chi}}^{(L)}_{MNP}\nabla^{0}\hat{\chi}^{MNP}-{17\over 18}\left(\dot{\hat{\chi}}_{M}\nabla^{0}\chi^{(L)M}+{\hat{\chi}}^{(L)}_{M}\nabla^{0}\hat{\chi}^{M}\right)\right] \nonumber \\
&& -{6\over\ell}\int d^2 x ~ \sqrt{-g}~ \varepsilon^{0RM}~ \dot{\hat{\chi}}_{MNP}\hat{\chi}_{R}^{~NP} +{2\over \ell}\int d^2 x ~ \sqrt{-g}~ \varepsilon^{0RM}\dot{\hat{\chi}}_{M}\hat{\chi}_{R} \nonumber \\
&& -{18\over \ell} \int d^2 x ~ \sqrt{-g}~ \varepsilon^{0RM}\dot{\hat{\chi}}_{MNP}{\chi}_{R}^{(L)~NP}+{17\over 3l} \int d^2 x ~ \sqrt{-g}~ \varepsilon^{0RM}~ \dot{\hat{\chi}}_{M}{\chi}^{(L)}_{R} \nonumber \\
E^{0}_{({\log}_{\Sigma})}&=& \int d^2 x ~ {\sqrt{-g}\over 2}~ \left( \dot{\hat{\Sigma}}_{MNP}\nabla^{0}\Sigma^{(L)MNP} +\dot{{\Sigma}}_{MNP}^{(L)}\nabla^{0}\hat{\Sigma}^{MNP} \right) \nonumber \\
&& -{2\over \ell}\int d^2 x ~ \sqrt{-g}~ \varepsilon^{0RM}~\dot{\hat{\Sigma}}_{MNP}\Sigma_{R}^{(L)NP}
\een

 All the integrands above are $t$ and $\phi$ independent. From the above expressions, one can easily see that for $M_{\Sigma}$, $L_{\Sigma}$ and $R_{\Sigma}$, the expression is quite simple, being given by single integrals, and by putting the solutions in the integrals, one find that they are negative. Hence one finds that $E^{0}_{R_{\Sigma}}$ is always positive, $E^{0}_{L_{\Sigma}}$ is positive for $\mu\ell >1$ and $E^{0}_{M_{\Sigma}}$ is positive for $\mu\ell<1$. And since $E^{1}$ vanishes for non-logarithmic modes, we find, in agreement with \cite{Chen:2011vp}, that the qualitative feature for the non-logarithmic $\Sigma$ modes is the same as that of the spin-2 case \cite{Li:2008dq}. The energy expressions for the left and right $\chi$ modes are obtained after putting the solutions in Mathematica as
\ben{E7}
E_{(L_{\chi})} &=& E^{0}_{(L_{\chi})}= {\pi\over 3\mu \ell^{6}}\left(1-\mu\ell\right), \nonumber \\
E_{(R_{\chi})} &=& E^{0}_{(R_{\chi})}={\pi\over 3\mu\ell^{6}}\left(1+\mu\ell\right).
\een
Thus we see that even for the $\chi$ modes the energy of the right branch is always positive and the energy of the left branch is positive for $\mu\ell <1$ and is zero for $\mu\ell=1$. Although a direct analytic expression for $E_{M_{\chi}}$ is not possible, but using Mathematica it can be seen that it is zero for $\mu\ell=1$, positive for $\mu\ell >1$ and negative for $\mu\ell <1$. We mention some of the numerical results for $E_{M_{\chi}}$ obtained using Mathematica.
\ben{E8}
\mu\ell&=&{1\over 3}: \quad  E_{(M_{\chi})}= E^{0}_{(M_{\chi})}=-{16\pi\over 45 \ell^5}.\nonumber \\
\mu\ell&=&1: \quad  E_{(M_{\chi})}= E^{0}_{(M_{\chi})}=0, \nonumber \\
\mu\ell&=&2: \quad  E_{(M_{\chi})}= E^{0}_{(M_{\chi})}={\pi \over 40 \ell^5}, \nonumber \\
\mu\ell&=&3: \quad  E_{(M_{\chi})}= E^{0}_{(M_{\chi})}={16\pi\over 315 \ell^5}.
\een
The energies $E^{0}$ for the logarithmic branch solutions are obtained (after putting the solutions in Mathematica) as:
\ben{E9}
 E^{0}_{({\log}_{\chi})} &=& {859 \pi \over 504 \ell^5}, \nonumber \\
 E^{0}_{({\log}_{\Sigma})} &=& -{132 \pi \over 25 \ell^5}.
\een
This, along with \refb{E32}, shows that the $\log_{\chi}$ modes has positive energy and the $\log_{\Sigma}$ modes has negative energy.
\subsection{Residual gauge transformation} \label{gauge}
In this section, we will show that the massless branch solutions and massive branch solution at the chiral point (both the trace as well as traceless modes) can be removed by an appropriate choice of residual gauge transformation. But since the residual gauge parameters does not vanish at the boundary, the modes can be regarded as gauge equivalent to the vacuum only if they have vanishing energy. Hence, as per the calculations of the energies above, we will see that massive and left moving solution at the chiral point (both the trace as well as traceless mode) can be regarded as gauge equivalent to vacuum. The gauge transformation in terms of the variable $\tilde{\phi}_{MNP}$ \refb{4.3} is
\ben{R1}
\delta \tilde{\phi}_{MNP} &=& \nabla_{(M}\xi_{NP)}+{1\over 2}\nabla_{Q}\xi^{Q}_{(M}g_{NP)}, \nonumber \\
\delta \tilde{\phi}_M &=& {9\over 2}\nabla_{N}\xi^{N}_{~M}.
\een
We need to find the residual gauge transformation obeying the gauge condition \refb{4.3} and the auxiliary condition \refb{4.9} implied by the equation of motion. We find that the residual gauge transformation satisfying these properties is
\ben{R2}
&& \nabla^2 \xi_{MN}-{6\over \ell^2}\xi_{MN}={3\over 4}\nabla_{(M}\nabla_{Q}\xi^{Q}_{~N)}, \nonumber \\
&& \nabla_{M}\nabla_{N}\xi^{MN}=0.
\een
One can use the above equation to deduce the following equation for $\nabla_{M}\xi^{M}_{~N}$
\be{R3}
\nabla^2\left(\nabla_{M}\xi^{M}_{~N}\right)-{34\over \ell^2}\nabla_{M}\xi^{M}_{~N}=0.
\ee
We thus see that $\nabla_{M}\xi^{M}_{~N}$ satisfies the same equation as $\tilde{\phi}_{M}$ \refb{6.2} at the chiral point $\mu \ell=1$, obeying the same condition \refb{4.9}. Thus one can choose the residual gauge transformation to remove the trace of the massless branch solution and of the massive branch solution at the chiral point which subsequently gauge away the appropriate $\chi$ modes. 

For the traceless $\Sigma$ modes, the residual gauge transformation should obey the equations
\ben{R4}
\nabla^2 \xi_{MN}-{6\over \ell^2}\xi_{MN} &= & 0, \nonumber \\
\nabla_{M}\xi^{M}_{N} &=& 0.
\een
We can once again see from \refb{R4} that for the residual gauge transformation parameter for the $\Sigma$ mode satisfying the above equation \refb{R4}, $\nabla_{(M}\xi_{NP)}$ satisfies
\ben{R5}
\nabla^{2}\nabla_{(M}\xi_{NP)}&=& 0, \nonumber \\
\nabla^{M}\nabla_{(M}\xi_{NP)} &=& 0.
\een
These equations are the same as the massless $\Sigma$ equations of motion and massive equations of motion at the chiral point \refb{6.12} and $\Sigma$ gauge condition \refb{6.15} and hence one can appropriately choose the parameters to gauge away the massless branch solution for $\Sigma_{MNP}$ and massive branch solution for $\Sigma_{MNP}$ at the chiral point.

To summarise, we find that both the massless $\chi$ and $\Sigma$ modes and their respective massive modes at the chiral point can be gauged away by an appropriate choice of residual gauge transformation parameters. Since the gauge transformation parameters do not vanish at the boundary, the modes can however be treated as gauge equivalent to vacuum only if they have vanishing energy. Hence, as per the energy calculations in the previous section, the left branch solution and massive branch solution at the chiral point can be regarded as gauge equivalent to vacuum. Since the logarithmic modes do not satisfy the same equations as their left moving partners, they cannot be regarded as pure gauge and are therefore physical propagating modes in the bulk. Thus the logarithmic traceless modes indicate a genuine instability in the bulk since they carry negative energy.

\section{Asymptotic Symmetries and the Chiral Point}\label{symm}
In our analysis of three dimensional gravity with spin three fields, we have seen that while solving the equations of motion for the linearised spin three, we find that there is a point where the basis for the solution becomes insufficient to describe it. This is the indication of the development of a logarithmic branch to the solution. This happens at a point where $\mu \ell =1$. This is the same point where the 
spin-two excitations develop a logarithmic branch and the central charge of the left moving Virasoro
algebra vanishes. 

Topological Massive Gravity at the chiral point was conjectured to be dual to a logarithmic conformal field theory with $c=0$. In our bulk analysis above, we have provided indications that a similar picture emerges when one includes the spin-three fields. To further our understanding of the symmetries of the boundary theory, let us look at the asymptotic symmetry structures.

\subsection{The $c=0$ confusion}

The asymptotic symmetry analysis for the theory with spin three fields in AdS (without the parity violating gravitational C-S term) was performed recently in \cite{Henneaux:2010xg, Campoleoni:2010zq}. The asymptotic symmetry algebra that was obtained was the classical $W_3$ algebra. 

\begin{eqnarray} \label{w3}
\big[ \  L_m, L_n \ \big] &=& (m-n) L_{m+n} + {c \over 12} m (m^2 - 1) \delta_{m+n, 0}  \\
\big[ \ L_m,  V_n \ \big] &=& (2m - n) W_{m+n} \nonumber \\
\big[ \ W_m,  W_n \ \big] &=& {c \over 360} m (m^2 -1) (m^2 - 4) \delta_{m+n,0} + {16 \over 5 c} (m-n) \Lambda_{m+n}
\nonumber \\
&+& (m-n) \Big( {1 \over 15} (m+n+2) (m+n+3) -{1 \over 6} (m+2) (n+2) \Big) L_{m+n}, \ \ \
\nonumber
\end{eqnarray}
where
\begin{eqnarray} 
\Lambda_m = \sum_{n=-\infty}^{+\infty}  L_{m-n} \ L_n \ . \label{non-lin}
\end{eqnarray}
sums quadratic nonlinear terms. Here the central charge for both the Virasoro and the pure $W_3$ is given by the Brown-Henneaux central term $c= {3 \ell \over 2G}$ for AdS. 

When one adds the parity violating gravitational C-S term, in the case of the usual $AdS_3$ without any higher spin terms, one ends up with corrected central terms where the left-right symmetry is broken, viz. $c_{\pm} = {3 \ell \over 2G} (1 \mp {1\over \mu \ell})$. The ``chiral-point" corresponds to 
$\mu \ell =1$ where $c_{+} =0$. 

The shift of the central terms, which is the effect of gravitational anomalies on the boundary stress tensor \cite{Kraus:2005zm,Solodukhin:2005ah}, does not change with the addition of the spin three fields. Thus the asymptotic symmetry algebra for the bulk theory with the Chern-Simons terms added is two copies of $W_3$ algebra, now with differing central charges. 

Now, when we look at the chiral point of the $W_3$ algebra, we see a potential problem. The non-linear term \refb{non-lin} in \refb{w3} has a coefficient which is inversely proportional to the central term and hence in the chiral limit would blow up. 

\subsection{The solution}

We propose a simple solution to the above problem. The blowing up of an algebra in a particular limit is indicative of the fact that one should look at an In{\"o}n{\"u}-Wigner contraction of the algebra at that point. 
To achieve this, let us rescale the generators as follows:
\begin{equation} \label{re-label}
L_n = L_n, \quad Y_n = \sqrt c W_n. 
\end{equation}
The rescaled W3 algebra now looks like
\begin{eqnarray} 
\big[ \  L_m, L_n \ \big] &=& (m-n) L_{m+n} + {c \over 12} m (m^2 - 1) \delta_{m+n, 0},  \\
\big[ \ L_m,  Y_n \ \big] &=& (2m - n) Y_{m+n}, \nonumber \\
\big[ \ Y_m,  Y_n \ \big] &=& {c^2 \over 360} m (m^2 -1) (m^2 - 4) \delta_{m+n,0} + {16 \over 5 } (m-n) \Lambda_{m+n}
\nonumber \\
&+& c (m-n) \Big( {1 \over 15} (m+n+2) (m+n+3) -{1 \over 6} (m+2) (n+2) \Big) L_{m+n}. \ \ \
\nonumber
\end{eqnarray}

Now, at the chiral point, the algebra would be the contracted version of the W3 algebra.
\begin{eqnarray} 
\big[ \  L_m, L_n \ \big] &=& (m-n) L_{m+n}, \quad \big[ \ L_m,  Y_n \ \big] = (2m - n) Y_{m+n}, \\
\big[ \ Y_m,  Y_n \ \big] &=& {16 \over 5 } (m-n) \Lambda_{m+n}. \nonumber
\end{eqnarray}

The $Y$ and $\Lambda$ actually generate an ideal and so one must set them to zero in any irreducible representation of the $W_3$ algebra. So the classical $W_3$ in the chiral limit essentially reduces to the Virasoro algebra. 

What we are advocating here is the classical analogue of what happens for the quantum $W_3$ for $c=-22/5$ \cite{Watt-notes}. Let us remind the reader of the quantum version of the $W_3$ algebra is.
The quantum effects enter into the regularisation of the quadratic non-linear term \refb{non-lin}. This shifts the overall quadratic coefficient of the quadratic term from ${16\over{5c}} \to {16\over{5c +22}}$ in \refb{w3}. As is obvious, $c=-22/5$ represents a blowing up of the quantum $W_3$ algebra and \cite{Watt-notes} prescribes a similar procedure to what we have outlined above. 

The logarithmic degeneracy at the chiral point that we would go on to construct, in this light would be related to a left moving LCFT with $c=0$, very similar to the original construction of the spin-two example. 

\subsection{Comments on other possible solutions}

The above procedure is certainly a correct one, but one might think that this is not the most general 
procedure that can be followed at the chiral point. Let us comment on a couple of other possible solutions. 

One way to argue that $c=0$ is not a problem in this context is to say that in this limit one should actually be looking at the quantum version of the $W_3$, instead of the classical algebra. Then the shifting of the non-linear term described above would mean that the algebra is perfectly fine in the chiral limit. When $c$ is small, and the curvature of space-time is large, it may be more sensible to look at the quantum algebra. The question obviously would be how an asymptotic symmetry analysis would see the change from classical to quantum and this is far from obvious. That this feature does not have any analogue in the well-studied spin-two example makes this an attractive avenue of further exploration. 

Another possible solution is to say that nothing is wrong at $c=0$. $\Lambda$ is actually a null field and the c=0 singularity is cancelled by $\Lambda$ become null. 
Let us take the quantum counterpart $c=-22/5$. 
Let us suppose that $\Lambda$ is a null field. 
We can work the commutation relations and see for example,
\be{}
\big[ \ L_m,  \Lambda_n \ \big] = (3 m - n) \Lambda_{m+n} + {{22+5c}\over 16} [m(m^2-1) L_{m+n}].
\ee
So we see that indeed at $c=-22/5$, this commutator closes to $\Lambda$. This is consistent with the fact that $\Lambda$ is a null field. We can similarly work out the consequences for $W_n$. 
The obstacle in this path is trying to figure out how to carry out an essentially quantum mechanical analysis in a classical algebra. We leave these issues for future work. 

\section{Conclusions and Future directions}\label{conc}
In this paper, we reviewed the the linearised action for spin-3 Fronsdal fields with a Chern- Simons term in flat space \cite{Damour:1987vm} and generalised it to $AdS$ space. The structure of the action is uniquely fixed by gauge invariance. We looked at its relation to the $SL(3,R) \times SL(3,R)$ Chern-Simons action \cite{Campoleoni:2010zq, Henneaux:2010xg} with unequal levels and fixed the normalisation of the gauge invariant action found earlier. We then looked at the equations of motion and decomposed it into left, right and massive branch. 

We figured out that the trace cannot be set to zero unlike the spin-2 case \cite{Li:2008dq}. The trace gives rise to non-trivial solutions to the equations of motion which has no counterpart in the spin-2 case. The trace solution has a ``resonant" behaviour at $\mu \ell={1\over 2}$. The massive branch trace mode carries positive energy for $\mu \ell >1$ and negative energy for $\mu \ell<1$ and zero energy for $\mu \ell=1$. The left branch solution carries positive energy for $\mu \ell <1$ and negative energy for $\mu \ell>1$ and zero energy for $\mu \ell=1$. Apart from the ``trace" solutions we also have the usual traceless mode. However the traceless mode has energy behaviour which is opposite to that of the trace mode (and similar to the spin-2 counterpart \cite{Li:2008dq}) i.e massive traceless mode carries positive energy for $\mu \ell <1$ and negative energy for $\mu \ell>1$ and zero energy for $\mu \ell=1$  and the left branch traceless solution carries positive energy for $\mu \ell >1$ and negative energy for $\mu \ell<1$ and zero energy for $\mu \ell=1$. The right branch solution carries positive energy for both the trace and traceless mode. 

At the chiral point the massive and left branch solution coincide and develop a new logarithmic branch both for the trace and traceless modes. The logarithmic solution for the trace mode carries positive energy whereas the logarithmic solution for the traceless mode carries negative energy. We also found that left branch and massive branch solution at the chiral point are pure gauge and have vanishing energy and hence can be treated as gauge equivalent to the vacuum. But the logarithmic modes are not pure gauge and are therefore physical propagating modes in the bulk. And since the logarithmic solution for the traceless mode carries negative energy, it indicates an instability in the bulk at the chiral point. It is therefore tempting to conjecture that higher spin massive gravity constructed in this paper at the chiral point is dual to a higher spin extension of $LCFT_{2}$. But there are some conceptual issues which should be dealt with before making this conjecture which are:
\begin{enumerate}
\item
\underline{Variational principle is well defined for the new logarithmic solutions}:
\\
The logarithmic solutions are the non trivial solutions to spin-3 massive gravity at the chiral point that grows linearly in time and linearly in $\rho$ asymptotically. It is found to have finite time-independent negative energy. But before it can be accepted as a valid classical solution one must check that the variational principle is well defined, i.e. the boundary terms vanish on-shell for the logarithmic solutions. Similar questions for the spin-2 counterpart was asked with an affirmative answer in \cite{Grumiller:2008qz}. We would also like to do similar check for both of our logarithmic solutions and as a by product obtain the boundary currents dual to the logarithmic modes.
\item
\underline{Consistent boundary conditions for the logarithmic modes}:
\\
We should be able to find consistent set of boundary conditions which encompasses the new logarithmic solutions i.e. there are consistent set of boundary conditions for which the generator of the asymptotic symmetry group is finite. Similar questions for the spin-2 case was asked with an affirmative answer in \cite{Grumiller:2008es}. We would also like to perform similar analysis for our logarithmic branch solutions.
\item
\underline{Correlation function calculation}:
\\
We should be able to compute correlation function in the gravity side. This should put us in a position to compare them with boundary correlators expected from a higher spin extension of LCFT. Similar questions were addressed in \cite{Skenderis:2009nt, Grumiller:2009mw} for the spin-2 case. The comparison in that case was however with correlators in LCFT which is well known in the literature. To our knowledge there is no higher spin extension of LCFT in the literature so far\footnote{See however some very recent work \cite{Rasmussen:2011rc}.}. The correlation function calculations should open up interesting questions to be answered about the higher spin extension of LCFT.
\item
\underline{One loop partition function calculation}:
\\
To make the higher spin extension of LCFT dual to the theory constructed in this paper more concrete, one should also compute the one loop determinant of the Euclidean theory constructed in this paper using the heat kernel techniques of \cite{David:2009xg} (which was also applied to the massless higher spin theory in \cite{Gaberdiel:2010ar}). If the LCFT proposal is right, one should be able to show that there would be no holomorphic factorisation of the one loop partition function at the chiral point. The expectation is that we would learn something more about the higher spin extension of LCFT from the structure of the one loop partition function. Similar calculations were done for TMG without higher spin in \cite{Gaberdiel:2010xv} and for General Massive Gravity in \cite{Bertin:2011jk} and the authors found concrete evidence for an AdS/LCFT picture.  In a subsequent work, we have looked at doing a similar computation for the spin-3 version of TMG constructed in this paper and subsequently generalized it to arbitrary spins \cite{in progress}.  

\end{enumerate}

Apart from all the above issues, the boundary CFT needs to be understood better. For example, there is the peculiar ``resonant" behaviour found for the trace modes at $\mu \ell={1\over 2}$ which should show up even in the CFT.  Apart from that we find a positive energy propagating mode in the bulk at the chiral point, which is the logarithmic solution corresponding to the trace mode. This has no counterpart in the spin-2 example and we would like to understand what this means from the CFT perspective. We leave these issues for future work.

Before we conclude, let us pause to remind the reader of the essential differences between our work and the work mentioned in the introduction which we said had some overlap with ours \cite{Chen:2011vp}. The part of our work which overlaps with \cite{Chen:2011vp} is the analysis of the traceless mode. The novel feature of our work is the trace modes and their logarithmic partner. We find several non-trivial features of this trace mode which we have addressed in this paper. We find instability in the bulk by explicitly computing the energy of the logarithmic partner of the traceless modes and end by speculating a higher spin extension of LCFT dual to the theory constructed in this paper at the chiral point. We also have a different proposal for the asymptotic symmetry structure and its peculiarities at the chiral point.

\subsection*{Acknowledgements} 
It is our pleasure to acknowledge helpful discussions and correspondence with Matthias Gaberdiel, Rajesh Gopakumar, Jose Figueroa O'Farrill, Ashoke Sen, Dileep Jatkar, Joan Simon, Marika Taylor, Daniel Grumiller and Yuji Tachikawa. We would especially like to thank Rajesh Gopakumar, Daniel Grumiller and Marika Taylor for valuable comments on the manuscript. AB would like to thank ICTP (Trieste), Ecole Polytechnique (Paris) and DESY (Hamburg) for hospitality during the course of this work. SL and AS would like to thank ICTP (Trieste), CHEP, IISC (Bangalore) and ICTS (Bangalore) for hospitality during the course of this work.


\appendix
{\section*{\underline{Appendices}}

\section{Taking the isometry generator across symmetrised covariant derivatives}\label{iso}
In this appendix we give the proof of the statement that the isometry generator can be taken across symmetrised covariant derivatives. Let the isometry generator be 
\be{a1}
L_{\xi}=\xi^{M}\partial_{M},
\ee
where $\xi_{M}$ satisfies
\be{a2}
\nabla_{(M}\xi_{N)}=0.
\ee
This generator acts on tensors of rank $(r,s)$ as
\ben{a3}
&& {\hspace{-0.5cm}} L_{\xi} T^{M_1 M_2......M^r}_{N_1 N_2...N_s} = \xi^{M}\partial_{M}T^{M_1 M_2......M^r}_{N_1 N_2...N_s}-\partial_{Q}\xi^{M_1}T^{Q M_2......M^r}_{N_1 N_2...N_s}-\partial_{Q}\xi^{M_2}T^{M_1 Q......M^r}_{N_1 N_2...N_s}\cdots  \nonumber \\
&& {\hspace{0.5cm}}-\partial_{Q}\xi^{M_r}T^{M_1 M_2......Q}_{N_1 N_2...N_s} + \partial_{N_1}\xi^{Q}T^{M_1 M_2......M^r}_{Q N_2...N_s}\cdots +\partial_{N_s}\xi^{Q}T^{M_1 M_2......M^r}_{N_1 N_2...Q} \nonumber \\
&& ={\hspace{0.5cm}}\xi^{M}\nabla_{M}T^{M_1 M_2......M^r}_{N_1 N_2...N_s}-\nabla_{Q}\xi^{M_1}T^{Q M_2......M^r}_{N_1 N_2...N_s}-\nabla_{Q}\xi^{M_2}T^{M_1 Q......M^r}_{N_1 N_2...N_s}\cdots  \nonumber \\
&&{\hspace{0.5cm}} -\nabla_{Q}\xi^{M_r}T^{M_1 M_2......Q}_{N_1 N_2...N_s} +\nabla_{N_1}\xi^{Q}T^{M_1 M_2......M^r}_{Q N_2...N_s}\cdots +\nabla_{N_s}\xi^{Q}T^{M_1 M_2......M^r}_{N_1 N_2...Q}.
\een
In the last equality we have added and subtracted Christoffel connections to write the partial derivatives as covariant derivatives. Now let us apply \refb{a3} to a tensor of rank 1 and its covariant derivative
\ben{a4}
L_{\xi}\phi_{N} &=& \xi^{M}\nabla_{M}\phi_{N}+\left(\nabla_N \xi^M\right)\phi_{M} \nonumber \\
L_{\xi}\left(\nabla_{P}\phi_{N}\right) &=& \xi^{M}\nabla_{M}\nabla_{P}\phi_{N}+\left(\nabla_N \xi^M\right)\nabla_{P}\phi_{M}+\left(\nabla_P \xi^M\right)\nabla_{M}\phi_{N}.
\een
Taking a covariant derivative of the first expression in \refb{a4} and subtracting it from the second, we obtain after some algebra
\be{a5}
\nabla_{P}L_{\xi}\phi_{N}-L_{\xi}\left(\nabla_{P}\phi_{N}\right)={1\over \ell^2}\xi_{[N}\phi_{P]}-\phi_{M}\nabla^{M}\nabla_{P}\xi_{N}.
\ee
Therefore symmetrising the indices we get
\be{a6}
\nabla_{(P}L_{\xi}\phi_{N)}-L_{\xi}\left(\nabla_{(P}\phi_{N)}\right)=0.
\ee
Now let us define $T_{PN}\equiv \nabla_{(P}\phi_{N)}$. Performing the same analysis as before we obtain
\be{a7}
\nabla_{M}\left(L_{\xi}T_{PN}\right)-L_{\xi}\left(\nabla_{M}T_{PN}\right)={1\over \ell^2}\left[\xi_{[P}T_{M]N}+\xi_{[N}T_{M]P}\right]-T_{PQ}\nabla^{Q}\nabla_{M}\xi_{N}-T_{NQ}\nabla^{Q}\nabla_{M}\xi_{P}.
\ee
And hence once again symmetrising the indices we get
\be{a8}
\nabla_{(M}\left(L_{\xi}T_{PN)}\right)-L_{\xi}\left(\nabla_{(M}T_{PN)}\right)=0.
\ee
Combining this with \refb{a6}, we get
\be{a9}
\nabla_{(M}\nabla_{N}L_{\xi}\phi_{P)}-L_{\xi}\left(\nabla_{(M}\nabla_{N}\phi_{P)}\right)=0.
\ee
This is what we wanted to prove.

\section{Conjugate momenta of different modes} \label{conjug}
In this appendix, we list all the conjugate momenta of the different modes that we obtained from the equation of motion. The conjugate momenta of the first kind are
\ben{Conj}
\Pi^{(1)MNP}_{(M_{\chi})}&=& {\sqrt{-g}\over 2}\left[-\nabla^{0}\chi^{(M)MNP}+{17\over 18\times 3}\nabla^{0}\chi^{(M)(M}g^{NP)} \right. \nonumber \\
&& ~~~~~~ -\left. {2\over \mu}\left(3\mu^2-{1\over \ell^2}\right)\varepsilon^{0R(M}\chi_{R}^{(M)~NP)}+{1\over 9\mu}\left(17\mu^2-{5\over \ell^2}\right)\varepsilon^{0R(M}\chi^{(M)}_{R}g^{NP)}\right] , \nonumber \\
\Pi^{(1)MNP}_{(L_{\chi})}&=& {\sqrt{-g}\over 2}\left[-\left(2-{1\over \mu \ell}\right)\nabla^{0}\chi^{(L)MNP}+{17\over 18\times 3}\left(2-{1\over \mu \ell}\right)\nabla^{0}\chi^{(L)(M}g^{NP)} \right.  \nonumber \\
&& ~~~~~~  -\left. {4\over \mu \ell^2}\varepsilon^{0R(M}\chi_{R}^{(L)~NP)}+{4\over 3\mu \ell^2}\varepsilon^{0R(M}\chi^{(L)}_{R}g^{NP)}\right] , \nonumber \\
\Pi^{(1)MNP}_{(R_{\chi})}&=& {\sqrt{-g}\over 2}\left[-\left(2+{1\over \mu \ell}\right)\nabla^{0}\chi^{(R)MNP}+{17\over 18\times 3}\left(2+{1\over \mu \ell}\right)\nabla^{0}\chi^{(R)(M}g^{NP)} \right.  \nonumber \\
&& ~~~~~~  -\left. {4\over \mu \ell^2}\varepsilon^{0R(M}\chi_{R}^{(R)~NP)}+{4\over 3\mu \ell^2}\varepsilon^{0R(M}\chi^{(R)}_{R}g^{NP)}\right] , \nonumber \\
\Pi^{(1)MNP}_{({\log}_{\chi})}&=& {\sqrt{-g}\over 2}\left[-\nabla^{0}\left[\hat{\chi}^{MNP}-\chi^{(L)MNP}\right]+{17\over 18\times 3}\nabla^{0}\left[\hat{\chi}^{(M}g^{NP)}-\chi^{(L)(M}g^{NP)}\right]\right.  \\
&& ~~~~~~  - \left. {4\over \ell}\varepsilon^{0R(M}\hat{\chi}_{R}^{~NP)}+{4\over 3l}\varepsilon^{0R(M}\hat{\chi}_{R}g^{NP)} - {12\over \ell} \varepsilon^{0R(M}{\chi}_{R}^{(L)~NP)}+{34\over 9l}\varepsilon^{0R(M}{\chi}^{(L)}_{R}g^{NP)}\right] . \nonumber 
\een
And
\ben{Conjj}
\Pi^{(1)MNP}_{(M_{\Sigma})} &=& {\sqrt{-g}\over 2}\left[-\nabla^{0}\Sigma^{(M)MNP}-{2\over 3\mu}\left(\mu^2 -{1\over \ell^2}\right)\varepsilon^{0R(M}\Sigma_{R}^{(M)NP)}\right], \nonumber \\
\Pi^{(1)MNP}_{(L_{\Sigma})} &=& -{\sqrt{-g}\over 2}\left(2-{1\over \mu \ell}\right)\nabla^{0}\Sigma^{(L)MNP} , \nonumber \\
\Pi^{(1)MNP}_{(R_{\Sigma})} &=& -{\sqrt{-g}\over 2}\left(2+{1\over \mu \ell}\right)\nabla^{0}\Sigma^{(R)MNP} , \nonumber \\
\Pi^{(1)MNP}_{({\log}_{\Sigma})} &=& {\sqrt{-g}\over 2}\left[-\nabla^{0}\left(\hat{\Sigma}^{MNP}-\Sigma^{(L)MNP}\right)-{4\over 3l}\varepsilon^{0R(M}\Sigma_{R}^{(L)NP)}\right].
\een
And the conjugate momenta of the second kind are
\ben{conj1}
\Pi^{(2)MNP}_{(M_{\chi})}&=& {\sqrt{-g}\over 2}\left[-g^{00}\chi^{(M)MNP}+{17\over 18 \times 3}g^{00}\chi^{(M)(M}g^{NP)}\right] , \nonumber \\
\Pi^{(2)MNP}_{(L_{\chi})}&=& {\sqrt{-g}\over 2}\left[-{1\over \mu \ell}g^{00}\chi^{(L)MNP}+{17\over 18 \times 3 \mu \ell}g^{00}\chi^{(L)(M}g^{NP)}\right]  , \nonumber \\
\Pi^{(2)MNP}_{(R_{\chi})}&=& {\sqrt{-g}\over 2}\left[{1\over \mu \ell}g^{00}\chi^{(R)MNP}-{17\over 18 \times 3 \mu \ell}g^{00}\chi^{(R)(M}g^{NP)}\right] , \nonumber \\
\Pi^{(2)MNP}_{({\log}_{\chi})}&=& {\sqrt{-g}\over 2}\left[-g^{00}\left[\hat{\chi}^{MNP}+{\chi}^{(L)MNP}\right]+{17\over 18 \times 3}g^{00}\left[\hat{\chi}^{(M}g^{NP)}+{\chi}^{(L)(M}g^{NP)}\right]\right]. \nonumber \\
\een
And
\ben{conj2}
\Pi^{(2)MNP}_{(M_{\Sigma})}&=& -{\sqrt{-g}\over 2}g^{00}\Sigma^{(M)MNP} , \nonumber \\
\Pi^{(2)MNP}_{(L_{\Sigma})}&=& -{\sqrt{-g}\over 2 \mu \ell}g^{00}\Sigma^{(L)MNP} , \nonumber \\
\Pi^{(2)MNP}_{(R_{\Sigma})}&=& {\sqrt{-g}\over 2 \mu \ell}g^{00}\Sigma^{(R)MNP} , \nonumber \\
\Pi^{(2)MNP}_{({\log}_{\Sigma})}&=& -{\sqrt{-g}\over 2}g^{00}\left[\hat{\Sigma}^{MNP}+\Sigma^{(L)MNP} \right].
\een
The labels $L$, $M$, $R$ and $\log$ labels labelling the left, massive, right and logarithmic modes respectively are kept inside ``( )" braces and hence should not be confused with the spacetime indices $MNP$. The following relations have been used
\ben{id}
&& \DD^{(L)}(\hat{\chi},\hat{\Sigma})_{MNP} \equiv (\hat{\chi},\hat{\Sigma})_{MNP}+{\ell\over 6}\varepsilon_{QR(M}\nabla^{Q}(\hat{\chi},\hat{\Sigma})^{R}_{~NP)}=-(\chi,\Sigma)^{(L)}_{MNP}, \nonumber \\
&& \DD^{(M)}{(\chi,\Sigma)}_{MNP}^{(M)}=\DD^{(L)}{(\chi,\Sigma)}_{MNP}^{(L)}=\DD^{(R)}{(\chi,\Sigma)}_{MNP}^{(R)}=0 ,\nonumber \\
&& \nabla^{2}\hat{\chi}_{MNP}={72\over \ell^2} \chi^{(L)}_{MNP}+{24\over \ell^2}\hat{\chi}_{MNP}+{2\over \ell^2}\hat{\chi}_{(M}g_{NP)} , \nonumber \\
&& \nabla^{2}{\chi}^{(L,R)}_{MNP}= {24\over \ell^2}{\chi}^{(L,R)}_{MNP}+{2\over \ell^2}{\chi}^{(L,R)}_{(M}g_{NP)}, \quad \nabla^{2}{\chi}^{(M)}_{MNP}=12\left(3\mu^2-{1\over \ell^2}\right)\chi^{(M)}_{MNP}+{2\over \ell^2}{\chi}^{(M)}_{(M}g_{NP)},  \nonumber \\
&& \nabla^{2}{\Sigma}^{(L,R)}_{MNP}=0, \quad \nabla^{2}{\Sigma}^{(M)}_{MNP}=\left(4\mu^2-{4\over \ell^2}\right) {\Sigma}^{(M)}_{MNP} , \nonumber \\
&& \nabla^2 \hat{\Sigma}_{MNP}= {8\over \ell^2}\Sigma^{(L)}_{MNP}.
\een


\end{document}